# Paradox in physics, the consistency of inconsistency (long version)

*DRAGOLJUB A. CUCIĆ*

*Regional centre for talents "Mihajlo Pupin", Pančevo, Serbia, cule@panet.rs*

Abstract

Paradoxes are a very frequent phenomenon in processes of thought which strive towards the intelectual and cognitive shifts. They occur in all areas of human spiritual activites. What we are interested here in, are the paradoxes in physics. We will try to give the answers to the questions such as: What are the reasons for developing the paradox in physics? Can a certain kind of classification of paradoxes in physics be nade and on what basis? What exactly is the paradox in phisics? What is the opinion of the physicists on paradox and what significance do they give to it?

*... but it is inconsistent and full of paradox.*

**Đorđe Živanović**

*To believe in it is logically possible without contradiction;*

*But it so strongly opposes my scientific instinct*

*that I cannot but search for a more complete conception.*

**A. Einstein**

*... that physicists have always learned the most from paradox.*

**R. Peierls**

## Introduction

Paradox is a frequent phenomenon present in every branch of human activity. It appears in art, science, religion and life. Paradox is, in itself, an anthropocentric phenomenon. It was created by man, who stands confused when faced with anything that is unexpected and different from his intuitive and rational understanding. Paradox represents the inconsistencies that do not fit the "world image".

A situation is considered paradoxical when, real or fictitious, it contradicts the general mode of thought. Paradox is used to show suspicion in the veracity of judgment. Through paradox a dilemma is presented that is to be solved by "correct reasoning". Paradox is a means of testing the correctness of thought.

Paradox is often a witty and very intelligent conclusion. Something other than expected is often found as the answer to a paradox. Thus we have, as an appropriate anecdote, the famous *Fermi's paradox* dealing with extraterrestrial intelligence. Enrico Fermi commented that *there probably are extraterrestrials in our galaxy, if the number of stars and the age of the Universe are taken into account. So, at least one of these civilisations is old enough and capable enough to spread over the entire galaxy. This should mean that the aliens are among us. The paradox is that we did not meet them yet.*

Similarly, the answer that does not respond to expectation is the one to the St. Augustine's paradox dealing with the measurement of time. St. Augustine concluded: *The past does not exist since it is no more, the future does not exist since it hasn't happened yet, and the present has no duration. The paradox is what we measure and how we do it.* These are beautiful examples of paradox as a game of intellect and a means to produce ambiguity.

Although it is an important and very intriguing (albeit somewhat infrequent) physical phenomenon, there are no studies that give answers to questions about paradox in physics. There is no collection point where paradoxes in physics can be found. Not even this study is it. And no texts can be found dealing with paradox as a physical phenomenon.

By analyses of different viewpoints in the understanding of paradox as an intellectual phenomenon and assertions that certain thought games and physical phenomena are essentially paradoxes, this text will attempt to build a consistent opinion on what paradox is, on the importance attributed to it by physicists and finally on what can be called a paradox in physics.

**Paradox**

The root of the present meaning of the word "paradox" can be found in the Latin *paradoxum* and Greek *paradoxos*,[1] which meant: contrary to opinion. Etymologically, a paradox is an apparent nonsense, since the logic of intellectual, expert public opinion is, in some way, contradictory to the paradox. In the middle ages, prior to the use of the word "paradox" in its present sense, the word *insolubilia* was used.

Thus, according to Svetislav Marić, a paradox is a statement contrary to expectation and general opinion. For S. Marić a paradox can be a new truth, but not necessarily – it can also be a falsehood used for fun as an intellectual game[2]. Vladimir Filipović is of a similar opinion as S. Marić, with the difference that he sees the paradox, unusual as it is, as a true statement (if examined thoroughly)[3]. Britannica tells us that paradox is an apparently self-contradictory statement, the underlying meaning of which is revealed only by careful scrutiny. In the Wikipedia, an online encyclopaedia, it can be found that: "*Paradox appears when a set of obviously irrefutable premises gives unacceptable or contradictory conclusions.*"

Aleksandar Bogojević considers paradox to be a truthful statement seemingly contrary to common sense. In a paradox everything happens "seemingly"; that is the conclusion drawn from suppositions that only seem to be true. For Aleksandar Bogojević a paradox is an error in our mode of thought.[4]

In *The Dictionary of Literary Terms* by Dragiša Živković it can be found that the word "paradox" means "unexpected" or "unusual" in Greek and was the name of lawsuits where the case was contrary to legal sentiment. At the schools of oratory, paradoxical cases were invented in order to help students develop their oratory skills. Also, in rhetoric, a paradox denotes concepts that contradict each other.

Stephen Barker distinguishes between three types of paradox:

In everyday language the term "paradox" is used to denote situations that appear impossible or contradictory but are still true (according to Kant's theory there is an equal number of even and odd numbers).

Paradox is the term for a proof that appears to be correct reasoning but leads to an absurd conclusion (paradoxes of Zeno).

Paradox is a situation in which, by seemingly correct reasoning, something can be shown to be both true and false (such a paradox is called an antinomy). A well known paradox is "Sophismata" or the paradox of Epimenides that states: Epimenides was a Cretan, Cretans always lie, all the other statements made by Cretans were false. [1]

When it is said that a certain conclusion contains a paradox it means that, according to the person who makes the judgment, the conclusion is incompatible with what is considered a universally acknowledged truth by public opinion of experts. This means that in making the conclusion some conditions were overlooked or the model used to make the conclusion is such that it favours conclusions that do not correspond to perception. A paradoxical situation is created when a conclusion is given as an answer to a question. A paradoxical situation produces confusion.

Based on what has been said a paradox could still be called an "apparent" incompatibility. This assertion is based upon the fact that paradox was formulated because a "proper" way to explain the causative situation had not been found. Paradox is caused by the lack of knowledge or by accepting an inappropriate model of knowledge.[5] It happens when one thing is expected and another thing received (while expecting one set of results, another appears). Paradox is an anthropomorphic phenomenon. It originates in man, since man is the one who creates and forms a paradoxical situation. A paradox is formulated when the causes of an event cannot be understood. A situation is termed paradoxical when the event, real or not, opposes some general mode of thought. Paradox leads to doubt in the veracity of judgment. It is a dilemma that needs to be solved. A paradox is an apparently contradictory conclusion that originates from ignorance. A paradox originates in collision of paradigms. Paradoxes appear in situations where the "founding spirit" has not been able to govern the facts and the relations between them, and where two or more paradigmatic orders are still mixing. The ideal ground for the growth of paradoxes is while the scientific explanations are still unclear and there is still perturbation in the overlapping of paradigmatic bases.

Paradox occurs in many fields of human action: philosophy, mathematics, logics (paradox of Epimenides, the village barber paradox, Russell's paradox), physics, psychology, economy and elsewhere. This text deals with paradox in physics.

**Paradox in physics**

Paradox in physics is, and is not, a frequent phenomenon, depending on the stadium of development of knowledge about it. Unlike in the other fields of human activity the paradox in physics needs to be solved since, if it remains unsolved, the physical explanation that contains the paradox is in question. A consistent physical explanation, generally accepted by the community of physicists does not "put up with" paradoxes. Paradoxes in physics appear for the very reason of bringing certain standpoints and explanations to absurdity and rejecting them as inconsistent. Paradoxes in physics are of critical importance since they set up standards of acceptance and rejection.

According to A. Bogojević paradoxes in physics are created "*when you confront the precise system of theory with common sense and intuition*"[6]. The last period of their expansion was the beginning of the 20th century due to the appearance of STR and GTR and quantum mechanics. His claim is that classical physics is "*a precise codification of our immediate experiences*" and that, during the development of STR, GTR and quantum mechanics, in the already formed paradigmatic setting of classical physics, a "conflict" occurred between intuitive expectations and the answers provided by the new theories. This resulted in the birth of many paradoxes: twins' paradox, Schrödinger's cat, EPR paradox, etc. As A. Bogojević himself stated: "*It turned out that former intuition in this domain failed us altogether. Solving the new formed paradoxes, we, in the end, developed a new intuition.*" According to A. Bogojević these theories were accepted only after the paradoxes had been solved. He considers paradox to be, as a phenomenon and subject matter, obsolete, since paradox no longer exists in modern science (namely physics). He has a high opinion of paradox as a phenomenon and considers it to be an instrument of cognition that contributed greatly to the development of 20th century physics.

Paradoxes did not originate only from relativistic and quantum-mechanical deliberations. In statistical physics we have the famous Gibson's paradox and the Maxwell's demon paradox. Other well known paradoxes are "reversible paradox" and "recurrent paradox" – based on Boltzmann's H-theorem. Cosmology, that contains the entire physics, had Olbers' and Zeliger's paradox.

Paradoxes in physics also appear in the form of fictitious experiments.[7] Both of them (paradox and fictitious experiment) are phenomena in physics, created, usually, for the purpose of intellectual provocation. Frequently, a paradox and a fictitious experiment are a single, inseparable, complementary phenomenon, serving, in the absence of technological skill to perform a real laboratory experiment, to pose fundamental questions, the answers to which build new paradigmatic bases in physics (EPR paradox, Schrödinger's cat paradox, twins' paradox ...). A fictitious experiment can contain a paradox, but doesn't need to. The existence of a paradoxical situation does not imply the existence of a fictitious experiment. A paradox, in its self, has nothing in common with a fictitious experiment, except that it is sometimes its integral part, as a form of expression. A paradox in a fictitious experiment implies that the result of the fictitious experiment defies "common sense" or the concept of some other epistemological model which interprets the result differently.

Certain branches of physics seem to be more "prone" to paradoxes than others. Some of these branches of physics where paradoxes appear more often are: quantum mechanics, theory of relativity and statistical physics. The reason why this happens is probably the very nature of these branches, which is more indirect than anywhere else. At the time of their creation they were not as "tangible", in terms of laboratory experimentation, as the more "accessible" ones: classical mechanics, thermodynamics, electrodynamics and optics.

To conclude with, let us return to the beginning. In the quoted sentence of Živanović's, in which he talks about the "classical" cosmological model with no intention of interpreting the meaning of paradox, he points out the essence of the cause of paradox in physics. But not only in physics, since it is the inconsistency of models and theories that brings about the possibility of paradox. A number of paradox examples will be presented in this text in order to confirm the aforementioned claims.

**Examples of paradox in physics, incomplete collection**

Olbers' paradox is a cosmological paradox, springing from the bases of the classical model of the universe across which the stars are spread out evenly. It poses the question: *Why is the sky, which should not be less bright than the stars (the Sun), black?*

The light of the stars fades as a consequence of absorption by interstellar matter that has not yet reached the state of thermodynamic equilibrium. When it reaches this state it will emit as much energy as it absorbed. Detection of the red shift in the light of distant galaxies eliminated the Olbers' paradox. The red shift indicates that the light brings less energy to the observer the more distant the star or galaxy is.

The "heat death" or Clausius' paradox is a thermodynamical paradox dealing with the fact that every physical system tends to the achievement of thermodynamic equilibrium. If the assumption is that the universe is eternal, a question arises: *How is it that thermodynamic equilibrium has not been long achieved?*

The scientific community of today rejects the assumption that the universe is eternal (Big Bang theory), so that thermodynamic equilibrium has not been achieved because there has not been enough time.

Zeliger's paradox is a cosmological paradox. According to the classical model of the universe, dominant in the 19th century, with the stars spread out evenly, gravitational potential is an undetermined value. *Universe affects every body in it with an undetermined force because of the large number of bodies it is made up of, that are evenly arranged around the observed body*.

Zeliger's paradox was eliminated by the Friedman's model of relativistic cosmology and the rejection of the stationary distributed matter; since Hubble's red shift showed that all the galaxies were moving away from each other.[8]

Twins' paradox is a relativistic paradox. *There is a pair of twin brothers. One of them is an astronaut and is to go on a space trip. The space ship the brother is going to fly on is able to reach near-light speed. The astronaut brother boards the space ship and sets off into space. After a while he returns. His twin brother, who stayed on Earth, is an old man*.

Due to dilatation of time, much less time had elapsed for the astronaut (this paradox is simultaneously a fictitious experiment).

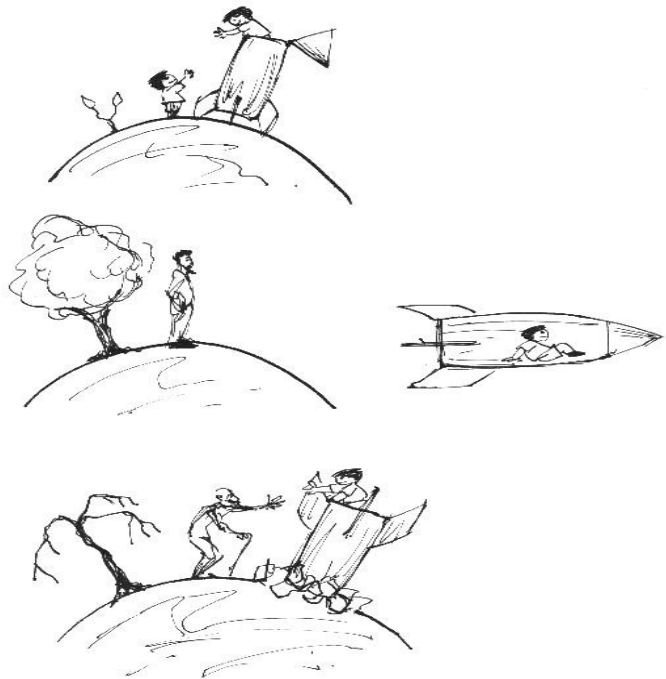

Picture 1. Twins' paradox[9]

The car and garage paradox is a relativistic paradox. *A man had a car and garage of equal proper length. The car stays in the garage and does not move until driven by the owner. One day he decided to try something. First he took the car outside and instructed the doorman to slam the garage door shut as soon as the back end of the car enters the garage. Then the driver speeds towards the garage.*

***According to the doorman***, *"the car contracted due to Lorentz contraction and easily fitted into the garage and then I slammed the door."*

***According to the driver***, *"the garage contracted due to Lorentz contraction and was too small for the car when I entered the garage."*

This is an example of a fictitious experiment that points out the relativity of the term "simultaneous". It is a paradox created by the relativistic effect of length contraction.

EPR paradox – the paradox of distant correlations is a quantum-mechanical paradox in which Einstein, Podolsky and Rosen consider the case of two sufficiently removed particles with a single wave function in order to demonstrate the existence of hidden variables. *By measuring one particle, the effect of the other particle will be established. For example, if the position of one particle is measured, the position of the other can be calculated. By measuring the impulse of one particle, the impulse of the other can be calculated. So, the position and the impulse of both particles can be measured. This is inconsistent with the assumption of the accuracy of Heisenberg's relation of indeterminacy.* [10]

A Swiss scientist, Bell, theoretically proved that the bases of the EPR paradox were not quantum-mechanical in character. Klauser-Friedman's laboratory experiment and, at the same time, A. Shimoni, and later A. Aspect all proved that hidden variables, that reduce the statistics of quantum mechanics to the causal principle, did not exist. The conclusion is that what the EPR paradox indicated did not fit the quantum-mechanical method of problem perception.

Schrödinger's cat paradox is a quantum-mechanical paradox. *Imagine an opaque box with: a radioactive source, a Geiger-müller counter, a glass phial with poison, a radioactive trigger and a live cat. The instrument in the box is set to crack the glass phial and kill the cat if the Geiger-müller counter registers a radioactive particle. If it does not, the cat will live.*

It is unknown to the observer whether the cat is alive or not. It is a question of statistics when the poison will be released from the phial due to radioactive decay. The observer does not know how long the cat will live. The fact is that the observer has placed the cat in the box, together with the poison phial and the radioactive trigger, and that he is unable to define the objective reality within the box from that moment on. The following conclusions are possible: the cat is alive, the cat is dead, the cat is both dead and alive.

Ultraviolet catastrophe paradox is one of few paradoxes in electrodynamics. *According to Rayleigh-Jeans formula, that explained radiation, for the wave lengths of ultraviolet radiation (in shorter wavelength* *radiation) an enormous amount of energy should be obtained, that would incinerate everything on Earth, since such is the radiation of the Sun.*

Since the Earth had obviously not been "incinerated" this paradox implied that something should be changed in the formula and the understanding of the problem. The solution came from a German scientist, M. Planck, who set up new bases for the problem, finally resulting in Planck's radiation law and the creation of the term "quant". From this, not of Planck's will, quantum mechanics was born.

Time symmetry paradox. Those who know about physics are aware that time is symmetrical and that many laws of physics are based on this invariance, since this symmetry is accepted as one of the most fundamental principles in physics. *The paradox is why the entire nature is not invariant. Why is the direction of time so clear and asymmetrical in this "real classical world of ours"? Why are our lives directed in only one direction? Why can the things that happened never more come to pass?*

Maxwell's demon paradox is a paradox of statistical physics. *"Imagine a creature capable of following every single molecule along its path. Such creature, whose characteristics would basically be as final as our own, would be capable of something that we are not. This is because the molecules in an air-filled container, at a regular temperature, move at velocities that are not regular at all, although the average velocity of a large, randomly selected, number of molecules is almost completely regular. Assume that such a container would be divided into two compartments, A and B, by a screen with a small hole in it. The creature that can see individual molecules would open and close the hole so that only the faster molecules may pass from A to B and only the slower from B to A. Thus, the temperature in B would increase and the temperature in A would diminish without work, which contradicts the Second law of thermodynamics"*. [8][11]

This is the original form of *Maxwell's demon*. The paradox is clear, since the demon (as the creature would be named later), as a nonhuman creature, is, actually, a creature of classical physics that behaves so nonclassically in the micro-world, thus creating the paradox.

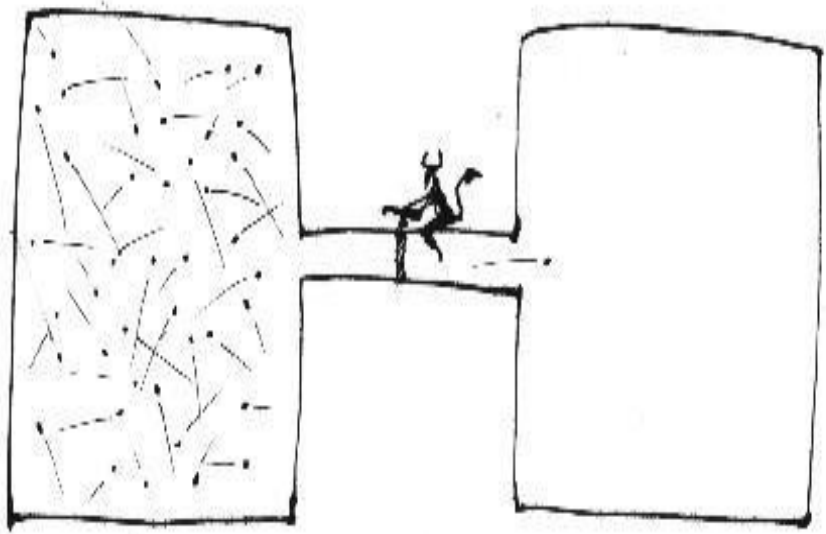

Picture 2. Maxwell's demon[12]

Gibbs' paradox is a paradox of statistical physics. This paradox occurs in the process of gas mixing. *With the removal of a screen and the mixing of two equal amounts of two equal gases (at an equal temperature, pressure and occupying equal volume) entropy increases. The paradox is that the removal of the screen is not a macroscopic process. The removal of the screen causes each of the two amounts of gas to spread through the entire volume and the system entropy, calculated according to the laws of classical statistics, changes, although, practically or physically speaking, nothing has changed, apart from the removal of the screen.*

Gibbs himself was the one who explained the paradox. Classical statistics assumes that identical particles may differ, while the states of a system resulting from permutation of identical particles may not. For this reason, classical statistics should be treated as a border case of quantum statistics. That is to say, the assumption should be made that identical particles may not differ.

The recurrent paradox is a paradox of statistical physics. This paradox is based on the Poincaré's theorem: *A system has final energy that is limited in final volume. After a sufficiently long time, the system returns to a state very similar to the original state. Poincaré's theorem indicates the periodic property of Boltzmann's H-function, and considering the H-theorem, $dH/dt \leq 0$, a paradox appears.*

Loschmidt's reversibility paradox is a paradox of statistical physics. Loschmidt's comment on Boltzmann's H-theorem[13] derivative is: *If a gas is in a state of imbalance, with a nonmaxwellian velocity distribution, described by Boltzmann's H-function, the gas will in time reach Maxwell's distribution, declining according to Boltzmann's H-theorem. If the velocity directions of all the particles in the state of equilibrium are changed, the second Boltzmann's function H` of equal probability will correspond to the new state. By rotating the velocity direction of molecules, the gas should pass through states of increasing character, which contradicts Boltzmann's H-theorem.*

The solution of the paradox was given even before the paradox was formulated, by Lord Kelvin, who, while considering other things, came to the conclusion that gas molecules, after the H` state, revert to a series of H states, and so, in a declining H-function, reach the state of equilibrium.

Boltzmann's paradox is a paradox of statistical physics. It is a paradox formulated in the question: *How is it that reversible microscopic movement of atoms leads to irreversible macroscopic phenomena*?

Boltzmann claims that the solution of the paradox has nothing to do with any particularities of irreversibility of microscopic movement. Macroscopic systems are irreversible (a glass that falls off a table and breaks, will not come together and return to the table, as it would appear if played in reverse on video or CD). The nature is arranged in a way that makes microscopic systems reversible and macroscopic systems irreversible.

## Conclusion

The oxymoron used in the title has multiple meanings. An oxymoron is, in itself, a syntactic connection of words of opposite meaning, in the form of rhetoric paradox. Simultaneously, for the existence of paradox in physics, the form of inconsistency, which creates it, is necessary. Just such a way of thinking leads to consistency. When the "firmness of the soft" is achieved, knowledge is established. When the inconsistencies become consistent it is possible to say that a certain theory rises to the level of fact.

Paradox in physics, as long as physics is developing, will not become an anachronism. Paradox is a motivational factor of the "living" physics. Horgan's apocalyptic predictions of the end of physics make Man equal to God. It simply seems incredible that mankind has mastered Universe and the rules of physical reality, but that it is still so helpless when it comes to solving many sore problems of its existence. Horgan and his contemporaries themselves create a paradox with their claims, which they cannot adequately solve.

It seems that there are two ways to explain the causes of the shortage of paradoxes in modern physics. The first points to the fact that physics has become a complete, canonised, consistent whole, in which there is no more room for paradox. According to the second, physics is at an optimal level of knowledge for the present human capabilities, awaiting new breakthroughs that will come when the conditions are right. This means that there will be more paradoxes that will still be one of the methods of stabilisation. Whatever the answer, and the right one will show itself, a paradox remains an intellectual challenge, inseparable from physics, and it will always be present as an integral part of human being in its subjective encounter with nature.

### Acknowledgements

I wish to express my gratitude to Zoran Males academic painter and sculptor and to professor Ivan Anicin.

**Appendix**

The appendix is comprised of a table of paradoxes according to branches of physics. This classification emphasizes the frequency of paradox in physics.[14] Although somewhat imprecise, this classification, that includes a list of paradoxes that is certainly not final, compels us to think why paradoxes are more frequent in some branches than in others. Special attention should be paid to the stage in the development of a branch of physics at which paradoxes appear (which was omitted in the table).[15]

| Branch | Paradox |
|---|---|
| *Statistical physics* | Boltzmann's paradox |
| | Maxwell's demon |
| | Recurrent paradox |
| | Loschmidt's reversibility paradox |
| | Gibbs' paradox |
| *Relativistic physics* | Twins paradox |
| | Event horizon astronauts |
| | Car and garage paradox |
| *Quantum mechanics* | EPR paradox |
| | De Broglie's box paradox |
| | Schrödinger's cat paradox |
| *Electrodynamics* | Ultraviolet catastrophe paradox |
| *Cosmology* | Olbers' paradox |
| | Zeliger's paradox |
| | GZK paradox |
| *Fluid dynamics* | D'Alembert's paradox |
| *Physical mechanics* | Carol's paradox |
| *Thermodynamics* | Clausius' paradox |
| | Mpemba paradox |
| *Time* | Time symmetry paradox |
| | St. Augustine's paradox |
| | Time travel paradox |

Incomplete table of paradoxes according to branches of physics

[1] Paradox is a Greek compound made up of two elements *para* and *doxa*. *Para* is a Greek preposition with the meaning of opposition; such as: contrary to, therefore, against, error, mistake. *Doxa* is a noun with the meaning of fame, reputation, good name; figuratively, something gained. In our case the best translation would be opinion.

[2] Svetislav Marić, *Filozofski rečnik*, Dereta, Belgrade, 1991, p. 107.

[3] Vladimir Filipović, *Filozofski rečnik*, NZMH, Zagreb, 1989, p. 243.

[4] http://scl.phy.bg.ac.yu/popularization/paradox.ppt.

[5] When someone is in a paradoxical situation they realize that something contrary to their expectation is happening. This means that their assumptions, predictions, and the knowledge of what caused their confusion are, to a certain extent, erroneous (if event related). Also, a paradoxical situation can be caused by an erroneous analysis of the proposed situation.

[6] http://scl.phy.bg.ac.yu/popularization/paradox.ppt.

[7] M.A. theses: Dragoljub Cucić, *Fiktivni eksperiment u fizici* (Fictitious Experiment in Physics), 2001.

[8] Đorđe Živanović, *Problemi relativističke kosmologije (*Problems of Relativistic Cosmology*)*, ΘΕΩΡΙΑ, No. 14, Serbian Philosophical Society magazine, Belgrade, p. 41-47., 1979.

[9] Taken over from [9]. The drawing is by Zoran Maleš, a sculptor.

[10] Fedor Herbut, *Konceptualne osnove kvantne mehanike*, p. 24-25

[11] Taken over from J.C. Maxwell, *Theory of Heat*, Longmans; London, 1891, section "Limitations of the Second Law of Thermodynamics", p. 207-208

[12] Taken over from [9]. The drawing is by Zoran Maleš, a sculptor.

[13] If a function f(x, t) satisfies a definite form of differential equation and H is a definite integral in which the function figures, then $dH/dt \leq 0$.

[14] Time and cosmology are included in the branches although many would disagree with it; either because Time in itself is not a branch of physics or because it, itself, could be a scientific discipline like physics (temporology, as some authors call it).

[15] I think it would be worth investigating to determine the level of knowledge in a certain branch at the time paradoxes appeared.